\documentclass[10pt]{article}
\usepackage[a4paper,margin=1in]{geometry}
\usepackage{graphicx}
\usepackage{amsmath}
\usepackage{amssymb}
\usepackage{hyperref}

\usepackage{fancyhdr}
\usepackage{ifthen}

\usepackage{booktabs}
\usepackage{makecell}
\usepackage{multirow}

\usepackage{titlesec}
\usepackage{caption}

\title{Data-Driven Nonlinear Deformation Design of 3D-Printable Shells}

\author{Samuel Silverman$^1$\and Kelsey L. Snapp$^2$\and Keith A. Brown$^{2,3,4}$\and Emily Whiting$^1$}

\date{}

\newcommand\shorttitle{Data-Driven Nonlinear Deformation Design}
\newcommand\authors{Silverman et al.}

\begin{document}

\maketitle

\begin{abstract}
Designing and fabricating structures with specific mechanical properties requires understanding the intricate relationship between design parameters and performance.
Understanding the design-performance relationship becomes increasingly complicated for nonlinear deformations.
Though successful at modeling elastic deformations, simulation-based techniques struggle to model large elastoplastic deformations exhibiting plasticity and densification.
We propose a neural network trained on experimental data to learn the design-performance relationship between 3D-printable shells and their compressive force-displacement behavior.
Trained on thousands of physical experiments, our network aids in both forward and inverse design to generate shells exhibiting desired elastoplastic and hyperelastic deformations.
We validate a subset of generated designs through fabrication and testing.
Furthermore, we demonstrate the network's inverse design efficacy in generating custom shells for several applications.
\end{abstract}

\textbf{Keywords:} additive manufacturing, neural networks, inverse design, elastoplasticity, hyperelasticity

\footnotetext[1]{Department of Computer Science, Boston University, Boston, MA, USA.}
\footnotetext[2]{Department of Mechanical Engineering, Boston University, Boston, MA, USA.}
\footnotetext[3]{Division of Materials Science \& Engineering, Boston University, Boston, MA, USA.}
\footnotetext[4]{Physics Department, Boston University, Boston, MA, USA.}

\section{Introduction}

Additive manufacturing has unlocked the ability to create structures with complex geometries and customized mechanical properties.
These fabricated structures can be designed to exhibit unique stiffness variations\cite{Panetta2015, Martinez2019} and energy-absorbing capabilities\cite{Bates2016, Gongora2020}.
However, achieving specific mechanical behaviors, especially for large deformations involving significant plasticity and densification, demands a deep comprehension of the intricate relationship between design parameters and performance.
Gaining insights through manual iterative design and testing often proves impractical and leads to costly and time-consuming design cycles.
Researchers have employed self-driving labs\cite{Gongora2020, Erps2021, Snapp2024} to explore design spaces autonomously.
However, these systems are constrained by cost, complexity, and converging time.

Simulation techniques like the finite element method (FEM) and homogenization excel at modeling elasticity\cite{Panetta2015, Martinez2019, Bickel2010, Schumacher2015, Martinez2016} and fracture\cite{Yang2020, Li2024}.
However, such strategies often lose accuracy when representing large plastic deformations, impeding the design of structures with targeted elastoplastic behaviors.
Researchers have also developed plasticity simulations to achieve highly complex deformation behavior \cite{Li2022, Zong2023, Cirio2016}.
However, further testing is needed to evaluate these methods' ability to model the compressive behavior of thin shell structures as used in this study.
Consequently, we propose a neural network trained on experimental data to learn the design-performance relationship between 3D-printable shells and their compressive deformation behavior.

Forward design presents users with predicted performance, allowing them to manipulate designs to achieve desired behavior.
Data-driven approaches to predict mechanical behavior from material geometries have been applied in various fields, from composites\cite{Yang2020, Abueidda2019, Yang2021} to material microstructures\cite{BenChaabene2020, Herriott2020}.
However, iterative design loops with forward design are often ineffective due to the vastness of design spaces and the complexity of how individual design parameters affect performance.

On the contrary, inverse design directly identifies the designs that achieve a target performance goal.
Inverse design is inherently complex; one performance is likely achievable by numerous designs, making learning algorithm convergence difficult. 
This one-to-many challenge mirrors complexities from other disciplines, like inverse scattering\cite{Colton1998} and inverse kinematics problems\cite{Kucuk2006}.
Despite this increased complexity, inverse design empowers users to explore and generate designs with desired mechanical properties.

\begin{figure}[ht]
\centering
\includegraphics[width=0.78\columnwidth]{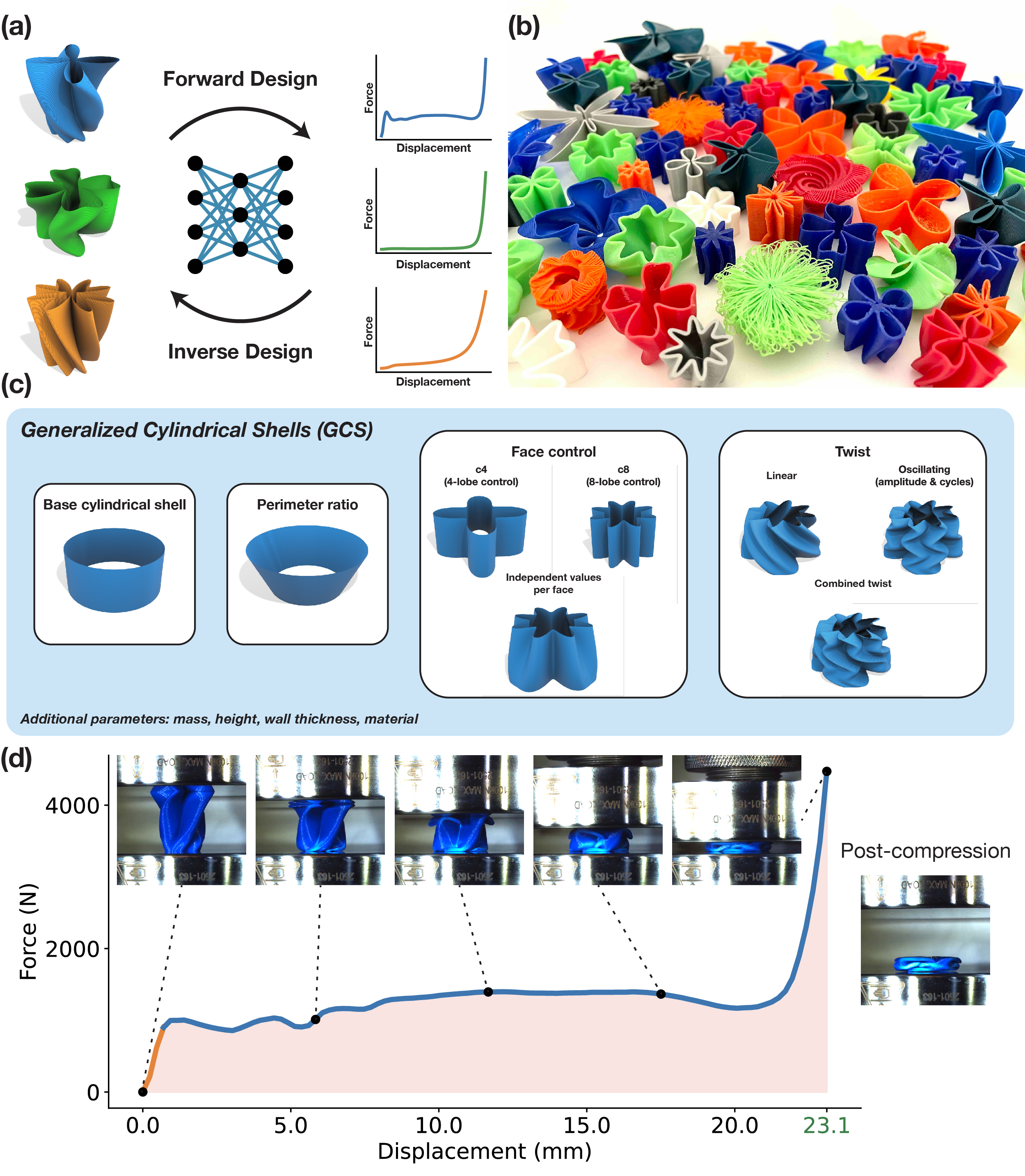}
\caption{\textbf{Overview}.
(\textbf{a}) We explore using a tandem neural network (TNN) for the forward and inverse design of generalized cylindrical shells (GCS).
(\textbf{b}) GCS are fabricated with fused deposition modeling (FDM) 3D printers.
(\textbf{c}) GCS geometry emerges from parameters that control a series of operations applied to cylindrical shells. The result of these operations is a diverse family of structures.
(\textbf{d}) Compression tests applied to GCS yield force-displacement curves\cite{Snapp2024}. Fabricated with PLA, this GCS exhibits elastoplastic deformation. Highlighted metrics are the linear elastic region used to calculate stiffness (orange line), work performed (red area under curve), and maximum displacement (green value).}
\label{fig:gcs}
\end{figure}

We propose a tandem neural network (TNN)\cite{Liu2018} for the forward and inverse design of a parametric family of cylindrical shells chosen for their ease of fabrication (Figure~\hyperref[fig:gcs]{\ref*{fig:gcs}a}).
The TNN combines two sequential neural networks: an inverse design network and a forward design network, structured like an autoencoder.
Researchers have used this architecture for the inverse design of nanophotonic devices\cite{Liu2018, Xu2021} and metamaterials\cite{Ma2018, Bastek2022, VantSant2023}.
Notably, machine learning-based inverse design extends beyond the TNN\cite{Deng2022, Li2023} with techniques ranging from convolutional neural networks\cite{Li2024} to reinforcement learning\cite{Gongora2021}.
Previous work generally focuses on mechanical properties that are easily modeled with simulation. 
These include properties arising from reversible elastic deformations or fracture propagation from an initial predetermined fracture site.

In this paper, we leverage an extensive experimental dataset comprising over 12,000 shells exhibiting nonlinear response to compression, as observed in their force-displacement curves, capturing a range of elastoplastic and hyperelastic deformations.
We verify our TNN's performance through experimentation on generated designs, compare the TNN to alternative methodologies, and demonstrate the TNN's effectiveness in generating designs with optimized nonlinear deformations through several applications, such as impact absorption.

\section{Materials and Methods}

This section explains our TNN pipeline.
We introduce the experimental dataset and preprocessing steps.
Furthermore, we describe the network architecture, learning objectives, and training process.

\subsection{Modeling Performance With Force-Displacement Curves}

We modeled performance with force-displacement curves (Figure~\hyperref[fig:gcs]{\ref*{fig:gcs}d}) and used the following derived metrics:

\begin{itemize}
    \item \textbf{Stiffness} (N/mm): Measures the resistance to initial deformations reflected in the slope of the linear elastic zone (orange region in Figure~\hyperref[fig:gcs]{\ref*{fig:gcs}d}).
    \item \textbf{Work} (J): Measures the energy absorption during deformation reflected in the area under the curve (red region in Figure~\hyperref[fig:gcs]{\ref*{fig:gcs}d}).
    \item \textbf{Maximum displacement} (mm): Denotes the furthest point of deformation reached during compression testing (green region in Figure~\hyperref[fig:gcs]{\ref*{fig:gcs}d}). It is influenced by material properties and experimental constraints. This value ensures a realistic scaling of displacements in predicted curves.
\end{itemize}

These metrics serve as high-level descriptors for evaluating model performance and selecting desired structures.
However, to ensure broad applicability, we maintain the entire force-displacement curve as the underlying performance representation, allowing users to identify metrics most pertinent to their unique design challenges.

\subsection{Generalized Cylindrical Shell Dataset}

In a comprehensive prior study, we conducted compression testing on 3D structures known as generalized cylindrical shells (GCS) (Figure~\hyperref[fig:gcs]{\ref*{fig:gcs}b}) to explore their energy-absorbing capabilities\cite{Snapp2024}.
These tests generated force-displacement curves, which, along with their associated designs, constitute a substantial GCS dataset\footnote{\url{https://hdl.handle.net/2144/46687}}.
This dataset holds particular value due to its wide range of measured elastoplastic and hyperelastic deformations.

Here, we provide an overview of the GCS parametric family but direct readers to the Methods section in our previous work \cite{Snapp2024} for a complete description.
The radius $r$ for an azimuthal angle $\phi$ defines a GCS's top and base faces,
\begin{equation}
    r(\phi) = r_0\big(1+c_4\cos(4\phi)+c_8\cos(8\phi)\big),
\end{equation}
where $c_4$ and $c_8$ control the shape and size of 4-lobe and 8-lobe features, respectively.
Interpolating between the two faces forms a cohesive 3D shell.
Parameters enable \textit{linear} and \textit{oscillating} twisting, enhancing geometry intricacy.
Moreover, adjustments to the \textit{perimeter ratio} between the top and base faces, \textit{mass}, \textit{height}, and \textit{wall thickness} allow further customization.
The $r_0$ term is a scale factor for perimeter size.
Figure~\hyperref[fig:gcs]{\ref*{fig:gcs}c} shows how a sequence of operations on cylindrical shells defines a GCS.

With a \textit{material} choice from one of six thermoplastics (two elastoplastic, three hyperelastic, one intermediate) the complete GCS specification requires 12 parameters (Table~\ref{tab:parameters}).
we used the MakerGear M3 and Ultimaker S5, two fused deposition modeling (FDM) 3D printers, to fabricate GCS by printing the design in vase mode.

\begin{table}[ht]
    \centering
    \caption{\textbf{GCS design parameters.}
    Twelve parameters define a GCS. We manually restricted the mass, height, and perimeter ratio values so that all parameters have well-defined ranges.}
    \label{tab:parameters}
    \begin{tabular}{lp{7.9cm}l}
        \toprule
        \textbf{Parameter} & \textbf{Description}  & \textbf{Dataset Values} \\
        \midrule
        $c_4$ (base \& top) & The parameter controlling the size and shape of the $4$-lobe feature. & $[0, 1.2]$ \\
        $c_8$ (base \& top) & The parameter controlling the size and shape of the $8$-lobe feature. & $[-1, 1]$ \\
        linear twist & The rotation (rad) of the top. This creates a linear twist between the base and top. & $[0, 2\pi]$ \\
        oscillating twist amplitude & The amplitude (rad) of the oscillating twist between the base and top. & $[0, \pi]$ \\
        oscillating twist cycles & The number of cycles of the oscillating twist between the base and top. & $[0, 3]$ \\
        perimeter ratio & The ratio between the top and base perimeters. & $[1, 3]$ \\
        mass & The mass (g). & $[1, 5]$ \\
        height & The height (mm). & $[10, 30]$ \\
        thickness & The wall thickness (mm). & $[0.4, 1]$ \\
        material & \makecell[tl]{Elastoplastic\\\hspace{0.5cm}PETG\\\hspace{0.5cm}PLA\\Hyperelastic\\\hspace{0.5cm}TPE (\textit{Chinchilla 75A})\\\hspace{0.5cm}TPU (\textit{Cheetah 95A}, \textit{NinjaFlex 85A})\\Intermediate\\\hspace{0.5cm}TPU (\textit{Armadillo 75D})} & Material choice \\
    \bottomrule
\end{tabular}
\end{table}

\subsection{Data Processing}

This section discusses our steps to extract the design and performance data from the GCS dataset, resulting in 12,706 design-performance pairs.

\subsubsection{Performance Dimensionality Reduction}

Force-displacement curves in the GCS dataset, which include thousands of points with varying spacings and displacement ranges, are impractical for prediction tasks.
We processed the curve data to 100 evenly spaced displacement values.
Inspired by Yang et al. \cite{Yang2020}, we used Principal Component Analysis (PCA) to condense the corresponding 100 force values into ten principal components.
Our performance vectors $\mathbf{p}\in\mathbb{R}^{11}$ combine these coefficients with the maximum displacement value.
Refer to supplementary \S1 for details on curve compression and analysis of PCA quality.

\subsubsection{Design Parameter Normalization}

Our investigation restricted the material parameter to experiments with at least 500 data points to ensure data quality.
For the GCS design parameters, we one-hot encoded the material parameter and applied min-max normalization to all non-material parameters to normalize their values.
However, the mass, height, and perimeter ratio parameters lack a clear range.
We manually capped these parameters' values to [1 g, 5 g] for mass, [10 mm, 30 mm] for height, and [1, 3] for perimeter ratio.

Using one-hot-encoded materials provided a straightforward way to ensure that the materials in generated designs conform to realistic values in inverse design.
Although we could have parameterized the material using a subset of continuous variables (e.g., Young's modulus and Poisson's ratio), we would have needed additional mechanisms to ensure realism in the generated values during inverse design.

These normalization operations collectively result in design vectors $\mathbf{d}\in\mathbb{R}^{17}$, comprising 11 geometric parameters and six values from the one-hot encoded material parameter.

\subsection{Model Architecture}

\begin{figure}[ht]
\centering
\includegraphics[width=0.5\columnwidth]{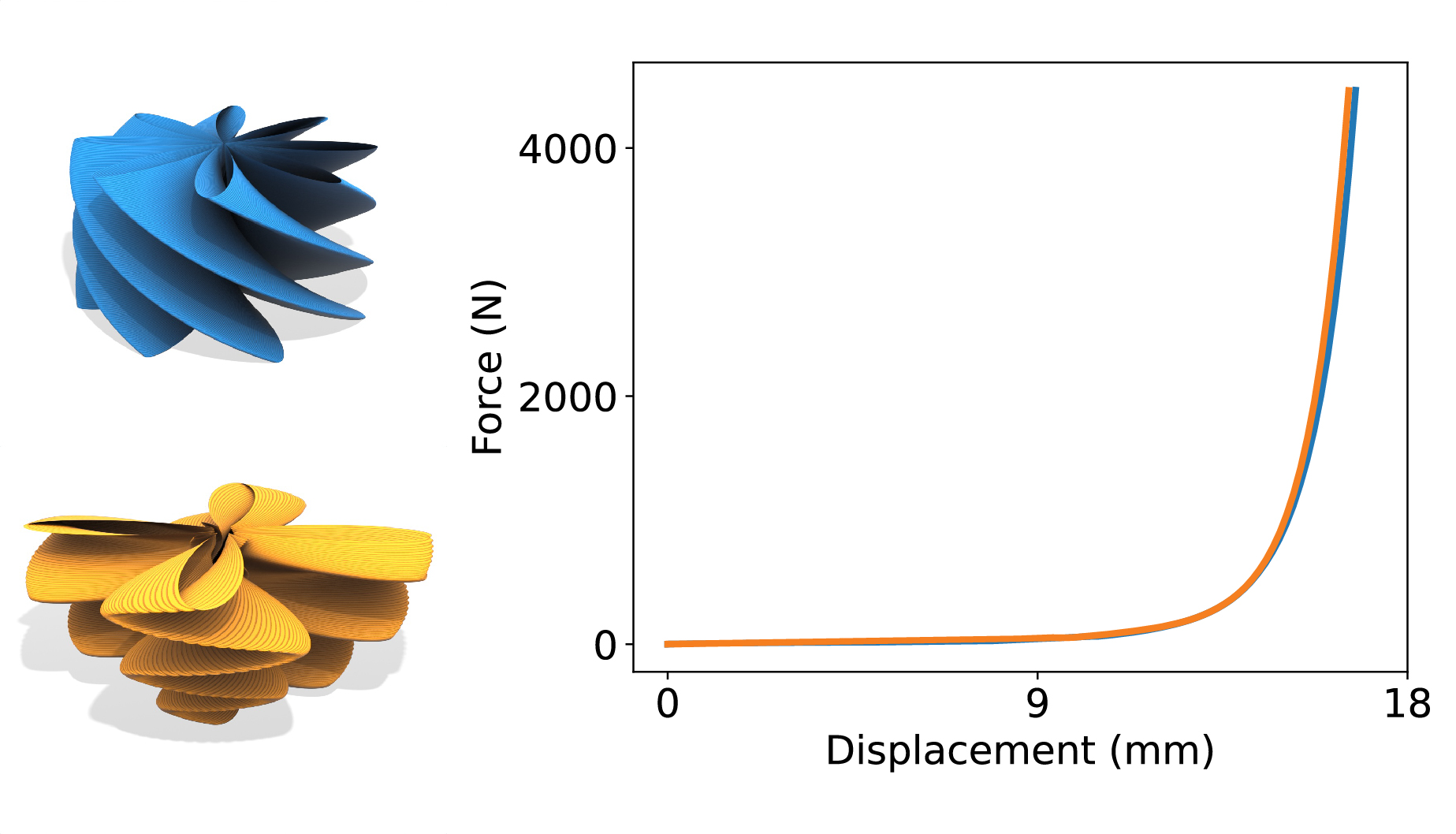}
\caption{\textbf{One-to-many performance to design relationship.}
Two GCS with distinctly different geometry share nearly identical force-displacement behavior, a common problem in inverse design.}
\label{fig:designspace}
\end{figure}

\begin{figure}[ht]
\centering
\includegraphics[width=0.75\columnwidth]{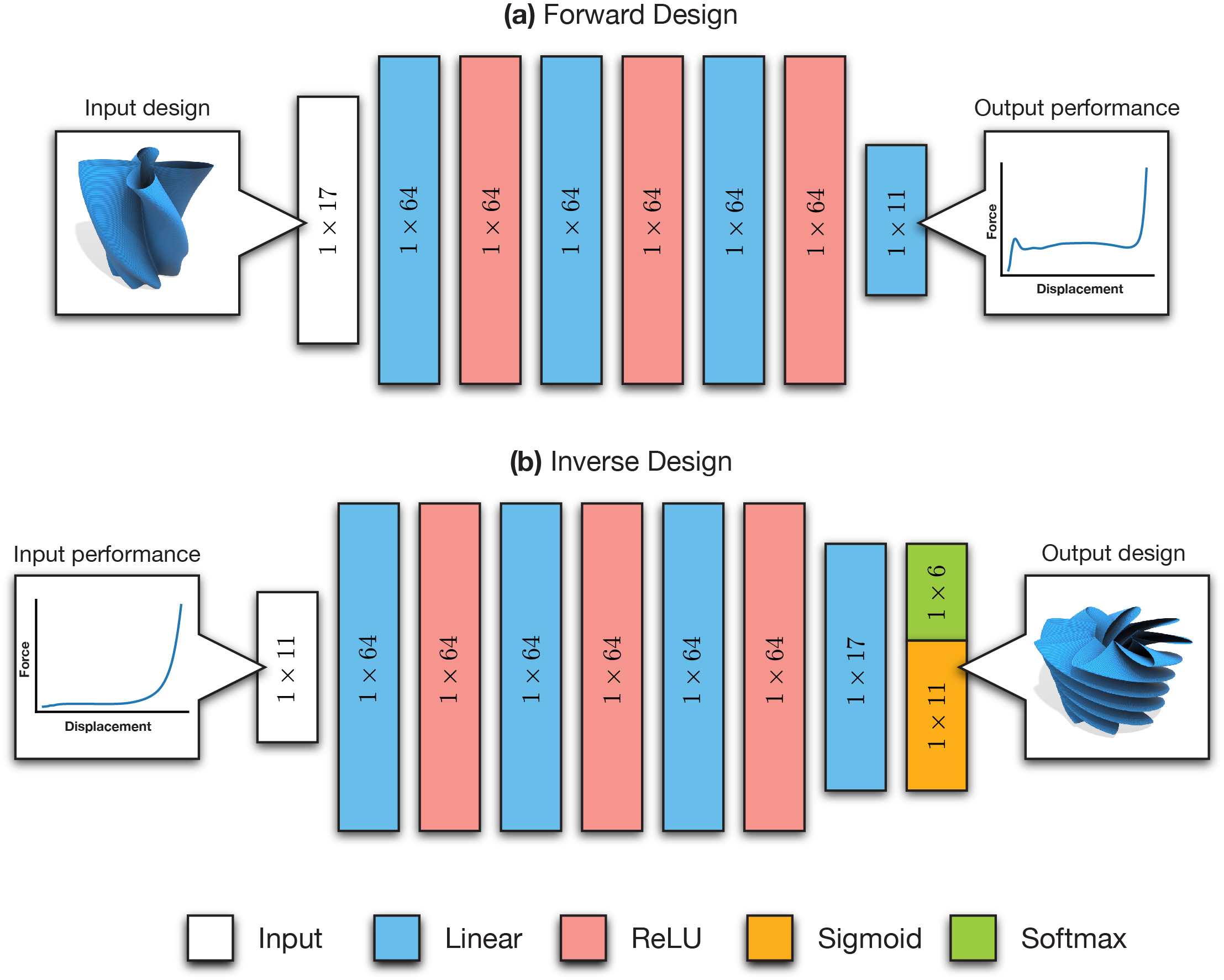}
\caption{\textbf{TNN architecture.}
    (\textbf{a}) The forward design network ($\mathcal{F}$) maps GCS designs to corresponding force-displacement curves.
    (\textbf{b}) The inverse design network ($\mathcal{I}$) maps force-displacement curves back to GCS designs.}
\label{fig:architecture}
\end{figure}

Inverse design poses complexity due to the potential one-to-many relationship between different designs and similar performances, hindering conventional learning algorithm convergence \cite{Liu2018}.
Figure~\ref{fig:designspace} depicts this challenge, illustrating two distinct GCS designs with nearly identical force-displacement curves.
To address this, the TNN framework, initially proposed by Liu et al. \cite{Liu2018}, has emerged as a promising solution.
We leveraged this framework to generate diverse GCS designs aligned with desired performances.
Figure~\ref{fig:architecture} shows our network architecture.

The forward design neural network $\mathcal{F}:\mathbb{R}^{17}\rightarrow\mathbb{R}^{11}$ learns the mapping from design vectors $\mathbf{d}$ to performance vectors $\mathbf{p}$.
The network architecture consists of six hidden layers: A linear layer followed by a ReLU activation, repeated three times.
All hidden layers maintain a uniform width of 64 units, contributing 10,190 trainable parameters.

The inverse design neural network $\mathcal{I}:\mathbb{R}^{11}\rightarrow\mathbb{R}^{17}$ learns the inverse mapping: performance vectors $\mathbf{p}$ to design vectors $\mathbf{d}$, and mirrors the architecture of $\mathcal{F}$.
The final layer of $\mathcal{I}$ ensures appropriate values for all GCS parameters by combining softmax and sigmoid activations.
The softmax activation is applied to the six parameters representing the one-hot-encoded material, assuring the predicted material has the highest value.
The sigmoid activation confines predicted values of the remaining parameters within $[0, 1]$, maintaining consistency with their normalized counterparts.

\subsection{Objective Function}

$\mathcal{F}$ aims to minimize the error between predicted and experimental performance vectors.
However, not all values in the performance vectors carry equal significance in improving the model's accuracy.
Building upon insights from Yang et al. \cite{Yang2020}, we used a weighted mean squared error (MSE) loss function, $\mathcal{L}_\mathcal{F}$, using a weight vector $\mathbf{w}\in\mathbb{R}^{11}$:
\begin{equation}
    \mathcal{L}_\mathcal{F}=\frac{1}{n}\sum_{i=1}^n\Big(\mathbf{w}\cdot\big(\mathbf{p}^{(i)}-\mathcal{F}(\mathbf{d}^{(i)})\big)\Big)^2,
\end{equation}
where $n$ is the number of samples and $\mathbf{w}$ is defined as
\begin{equation}
    \mathbf{w}_j=\left\{\begin{matrix}
        \frac{\lambda_j}{\sum_{k=1}^{10}\lambda_k} & \text{if }1\leq j\leq10 \\
        1 & \text{if }j=11 \\ 
    \end{matrix}\right..
\end{equation}
By setting $\mathbf{w}_{11}=1$, we assigned equal importance to the entries of the performance vector responsible for the displacement and force values.
However, since each principal component coefficient explains a different amount of variance in the force data, as indicated by the eigenvalues $\lambda_1,\ldots,\lambda_{10}$ obtained from PCA, we set the weights of individual coefficients based on their respective explained variance.
By incorporating these weighted values, $\mathcal{F}$ prioritizes the most informative principal component coefficients during training.

The objective of $\mathcal{I}$ is twofold.
First, $\mathcal{I}$ aims to generate designs that align with desired performances.
The loss function $\mathcal{L}_\mathbf{p}$ encourages this behavior, mirroring the form of $\mathcal{L}_\mathcal{F}$,
\begin{equation}
    \mathcal{L}_\mathbf{p}=\frac{1}{n}\sum_{i=1}^n\bigg(\mathbf{w}\cdot\Big(\mathbf{p}^{(i)}-\mathcal{F}\big(\mathcal{I}(\mathbf{p}^{(i)})\big)\Big)\bigg)^2.
\end{equation}
$\mathcal{L}_\mathbf{p}$ addresses the one-to-many mapping problem in inverse design by ensuring the generated GCS's predicted performance matches the target performance without directly considering the generated GCS design.
Second, $\mathcal{I}$ aims to generate printable designs.
Drawing inspiration from prior work \cite{Ma2018}, we used the loss function $\mathcal{L}_\mathbf{d}$ to bias generated designs towards previously tested designs,
\begin{equation}
    \mathcal{L}_\mathbf{d}=\frac{1}{n}\sum_{i=1}^n\big(\mathbf{d}^{(i)}-\mathcal{I}(\mathbf{p}^{(i)})\big)^2.
\end{equation}
Specifically, $\mathcal{L}_\mathbf{d}$ encourages predicted designs to align with the dataset designs associated with the target performances.
Such a loss is motivated by the fact that dataset designs are all known to be printable.
While it might be possible to include losses that penalize parameter combinations leading to non-printable designs, we do not explore these methods here.

$\mathcal{I}$ seeks to minimize the loss function $\mathcal{L}_\mathcal{I}$, which combines $\mathcal{L}_\mathbf{d}$ and $\mathcal{L}_\mathbf{p}$ in a weighted manner:
\begin{equation}
\mathcal{L}_\mathcal{I}=\mathcal{L}_\mathbf{p}+\alpha\mathcal{L}_\mathbf{d},
\end{equation}
where $\alpha\in\mathbb{R}$ determines the relative weight between $\mathcal{L}_\mathbf{p}$ and $\mathcal{L}_\mathbf{d}$.
By fine-tuning the $\alpha$ value, we can control the balance between optimizing for predicted performance accuracy and maintaining proximity to dataset designs.
While it may be possible to include losses that penalize parameter combinations that result in non-printable designs, we do not investigate such methods here.

\subsection{Training}
\label{sec:training}

We divided our processed GCS dataset into training, validation, and test sets, following an 80-10-10\% split.
Our training process involved two stages.
We trained $\mathcal{F}$ in the initial stage. 
Upon completion, we froze $\mathcal{F}$ and appended the untrained $\mathcal{I}$ model.
The second stage involved training $\mathcal{I}$ using $\mathcal{F}$ to aid in convergence.

Trained in this order, $\mathcal{I}$ learns to generate GCS designs whose performance predicted by $\mathcal{F}$ aligns with the desired performance.
Had we trained $\mathcal{I}$ independently, we could only use $\mathcal{L}_\mathbf{d}$ as $\mathcal{L}_\mathbf{p}$ depends on $\mathcal{F}$.
Alone, $\mathcal{L}_\mathbf{d}$ exposes an ill-posed learning problem, as multiple GCS designs could yield similar performances.
Penalizing deviations between generated and actual designs would impede convergence, rendering the learning process ineffective.
While it is likely possible to train both networks simultaneously, similar generator and discriminator training in generative adversarial networks \cite{Goodfellow2014}, we left this to future investigations.

We use the Adam optimizer \cite{Kingma2015} for training, with a learning rate of 0.001, a weight decay of 1, and a batch size of 16.
We used early stopping to prevent overfitting, terminating after 500 epochs in each stage.
Our TNN was implemented using  PyTorch \cite{Paszke2019} and trained on an Apple MacBook Pro (M1 Max).
Training both networks required less than an hour.

\section{Results}

We evaluated our TNN's forward and inverse design accuracy on the test set.
In our evaluation, we repeated training ten times using different random splits of the data to report test outcomes with 95\% confidence intervals.
We performed physical experimentation on a sample of generated GCS designs and compared the TNN performance to other methodologies.
Finally, we generated GCS with tailored mechanical properties for two applications to demonstrate our TNN's inverse design capabilities.

\subsection{Forward Design Performance}

We evaluated $\mathcal{F}$ on the test set by comparing the predicted and experimental force-displacement curves.
We compared the metrics of the predicted and experimental curves because using high-level metrics for evaluation provides interpretable units for measures such as mean absolute error (MAE).
Notably, the TNN did not directly learn metrics like work or stiffness; they are computed based on the learned performance vectors.

\begin{figure}[ht]
\centering
\includegraphics[width=\columnwidth]{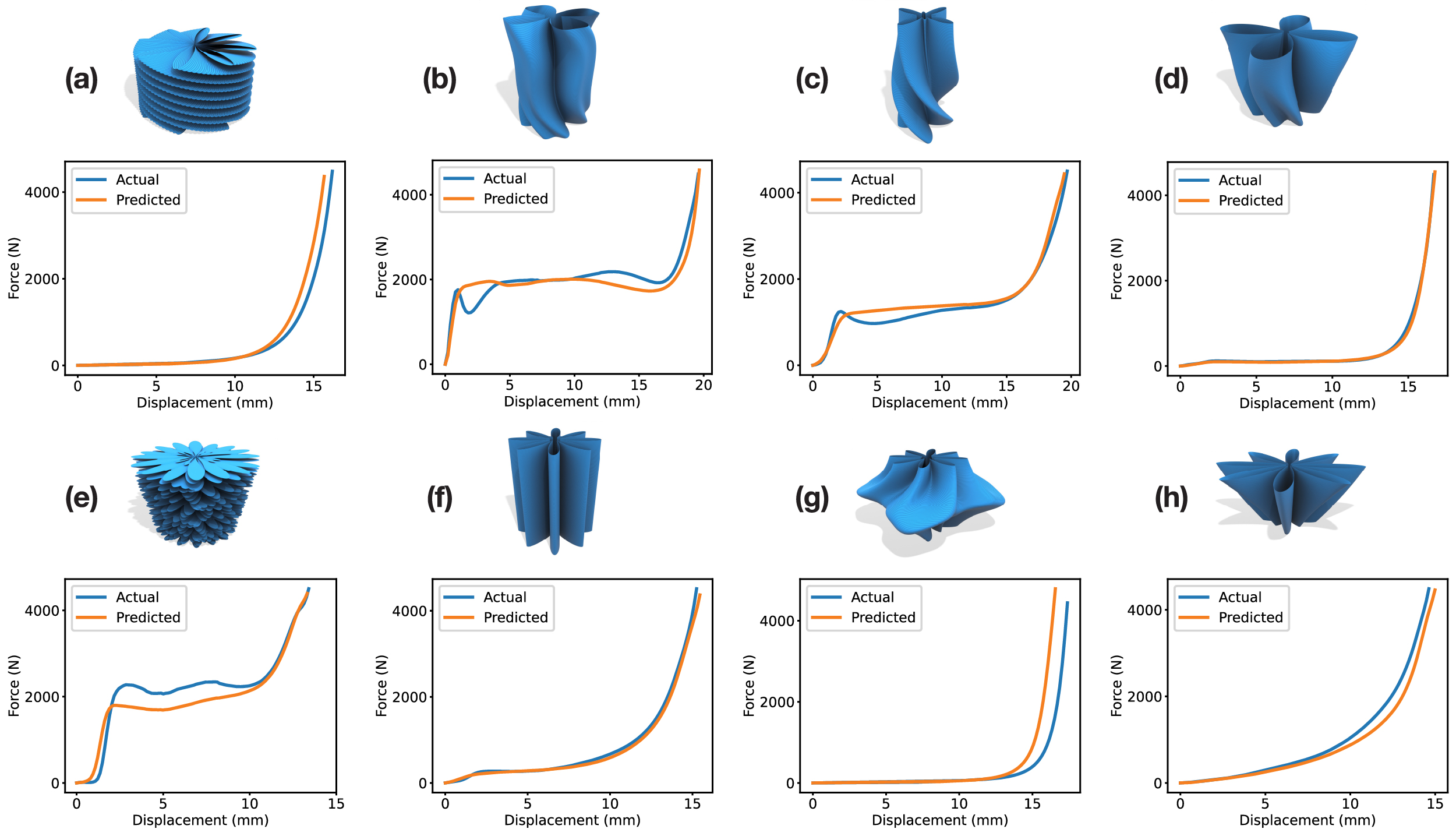}
\caption{\textbf{Forward design results.}
Eight randomly selected results from the test set.
The GCS designs (blue) serve as input to $\mathcal{F}$, which predicts force-displacement curves (orange). For reference, we show the actual experimental force-displacement curves from the test set (blue).
$\mathcal{F}$ can predict nonlinear deformation behavior for elastoplastic (\textbf{b}, \textbf{c}, \textbf{e}) and hyperelastic (\textbf{a}, \textbf{d}, \textbf{f}, \textbf{g}, \textbf{h}) GCS.}
\label{fig:forwardresults}
\end{figure}

Table~\ref{tab:forwardperformance} presents the MAE and R$^2$ for stiffness, work, and maximum displacement.
Each metric's MAE is $<5\%$ of their respective ranges: 3.2\% for stiffness, 2.8\% for work, and 2.2\% for maximum displacement.
Furthermore, the small confidence intervals demonstrate that our TNN has training stability, with minimal reliance on model initialization or data splitting.
Figure~\ref{fig:forwardresults} shows $\mathcal{F}$'s performance for eight test set GCS designs, demonstrating the network's capability to predict the complete nonlinear force-displacement behavior of designs made with elastoplastic and hyperelastic materials.
Refer to the supplementary material for the design parameters.

\begin{table}[ht]
\centering
    \caption{\textbf{Forward design performance.}
    The test set results show the work, stiffness, and maximum displacement of predicted force-displacement curves by $\mathcal{F}$.
    We report each metric's R$^2$ and mean absolute error (MAE) with 95\% confidence intervals.
    For reference, we provide the mean and range of each metric in the dataset.
    The test set loss is $\mathcal{L}_\mathcal{F}=0.21\pm0.01$.
    }
    \label{tab:forwardperformance}
    \begin{tabular}{lllll}
    \toprule
    & \multicolumn{1}{c}{\textbf{R}$^\mathbf{2}$} & \multicolumn{1}{c}{\textbf{MAE}} & \multicolumn{1}{c}{\textbf{Mean}} & \multicolumn{1}{c}{\textbf{Range}} \\
    \midrule
    Stiffness (N/mm) & $0.66\pm0.03$ & $250\pm20$ & 815 & 7732 \\
    Work (J) & $0.96\pm0.002$ & $1.3\pm0.04$ & 15.4 & 46.7 \\
    Max. displacement (mm) & $0.94\pm0.007$ & $0.50\pm0.03$ & 18.6 & 22.6 \\
    \bottomrule
\end{tabular}
\end{table}

\subsection{Inverse Design Performance}

We trained $\mathcal{I}$ with $\alpha\in\{0,0.01,0.1,1\}$, the parameter weighing the importance of generating previously tested designs.
To evaluate the inverse design performance, we compare predicted designs' predicted force-displacement curves $\mathcal{F}\big(\mathcal{I}(\mathbf{p})\big)$ to the target force-displacement curves $\mathbf{p}$.
Table~\ref{tab:inverseperformance} presents the MAE and R$^2$ for stiffness, work, and maximum displacement for $\mathcal{I}$ trained with each $\alpha$ value.

For $\alpha\in\{0,0.01,0.1\}$, we observed minimal change in accuracy for the work and maximum displacement.
However, for $\alpha=1$, we saw the MAE increase for maximum displacement and decrease for work.
For stiffness, the accuracy improved as we increased $\alpha$.
We found that different metrics derived from force-displacement curves exhibit varying sensitivity levels to the effect of $\mathcal{L}_\mathbf{d}$.
These findings underscore a complex trade-off between generating high-performing designs and printable designs.

\begin{table}[ht]
\centering
    \caption{\textbf{Inverse design performance.}
The test set results show the work, stiffness, and maximum displacement of predicted force-displacement curves by $\mathcal{I}$ trained on $\alpha\in\{0,0.01,0.1,1\}$.
We compare predicted designs' predicted force-displacement curves $\mathcal{F}\big(\mathcal{I}(\mathbf{p})\big)$ to the target force-displacement curves $\mathbf{p}$.
We report each metric's R$^2$ and MAE with 95\% confidence intervals. The test set loss is $\mathcal{L}_\mathcal{I}=0.0043\pm0.0005$ ($\alpha=0$), $\mathcal{L}_\mathcal{I}=0.0049\pm0.0005$ ($\alpha=0.01$), $\mathcal{L}_\mathcal{I}=0.0094\pm0.0003$ ($\alpha=0.1$), and $\mathcal{L}_\mathcal{I}=0.046\pm0.002$ ($\alpha=1$).}
    \label{tab:inverseperformance}
    \begin{tabular}{llll}
    \toprule
        & & \multicolumn{1}{c}{\textbf{R}$^2$} & \multicolumn{1}{c}{\textbf{MAE}} \\
    \midrule
    \multirow{3}{*}{$\alpha=0$} & Stiffness (N/mm) & $0.56\pm0.08$ & $310\pm30$ \\
    & Work (J) & $0.94\pm0.02$ & $1.8\pm0.3$ \\
    & Max. displacement (mm) & $0.99\pm0.002$ & $0.12\pm0.01$ \\
    \midrule
    \multirow{3}{*}{$\alpha=0.01$} & Stiffness (N/mm) & $0.60\pm0.06$ & $300\pm30$ \\
    & Work (J) & $0.92\pm0.04$ & $2.1\pm0.5$ \\
    & Max. displacement (mm) & $0.99\pm0.002$ & $0.12\pm0.01$ \\
    \midrule
    \multirow{3}{*}{$\alpha=0.1$} & Stiffness (N/mm) & $0.69\pm0.05$ & $250\pm30$ \\
    & Work (J) & $0.95\pm0.01$ & $1.7\pm0.2$ \\
    & Max. displacement (mm) & $0.99\pm0.002$ & $0.12\pm0.008$ \\
    \midrule
    \multirow{3}{*}{$\alpha=1$} & Stiffness (N/mm) & $0.68\pm0.04$ & $240\pm20$ \\
    & Work (J) & $0.98\pm0.004$ & $0.91\pm0.08$ \\
    & Max. displacement (mm) & $0.99\pm0.001$ & $0.23\pm0.01$
    \\
    \bottomrule
\end{tabular}
\end{table}

\subsubsection{Physical Validation}

\begin{figure}[ht]
\centering
\includegraphics[width=\columnwidth]{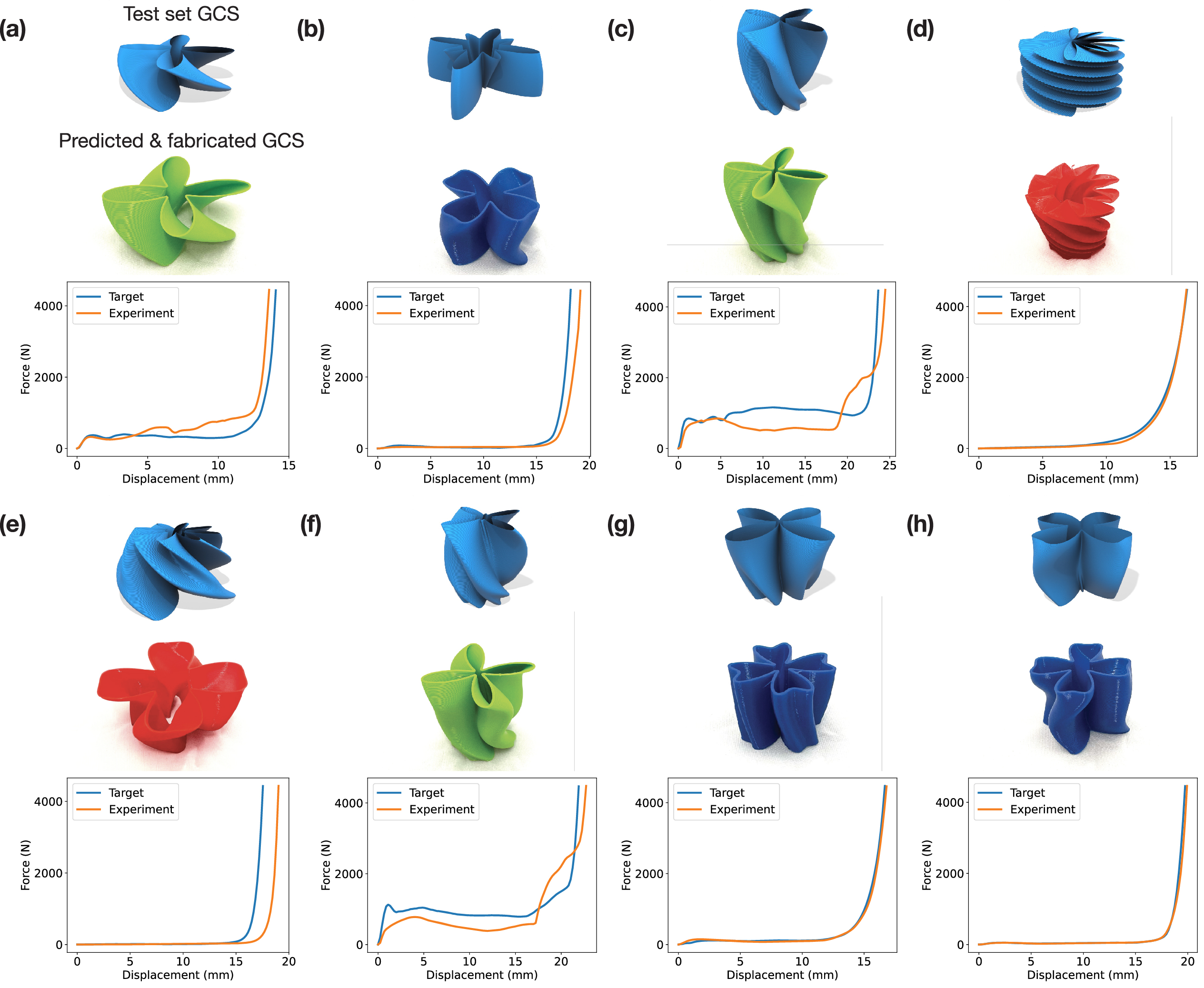}
\caption{\textbf{Inverse design results.}
Eight randomly selected results from the test set.
The force-displacement curves (blue) serve as input to $\mathcal{I}$, which predicts GCS designs.
We fabricated and performed compression testing on the predicted GCS designs to obtain experimental force-displacement curves (orange).
$\mathcal{I}$ can generate GCS designs that exhibit nonlinear elastoplastic (\textbf{a}, \textbf{c}, \textbf{f}) and hyperelastic (\textbf{b}, \textbf{d}, \textbf{e}, \textbf{g}, \textbf{h}) deformations.
For reference, we include the test set GCS designs (blue) associated with the inputted curves to illustrate the one-to-many relationship between performance and designs.
Generated designs differ, sometimes significantly, from their test set counterparts.}
\label{fig:inverseresults}
\end{figure}

We randomly selected eight GCS designs generated from $\mathcal{I}$ trained with $\alpha=1$ and experimentally obtained their force-displacement curves to assess the accuracy of predicted performance (Figure~\ref{fig:inverseresults}).
We fabricated the samples on a MakerGear M3 and performed compression testing with an Instron 5965.
Given a nonlinear force-displacement curve, $\mathcal{I}$ can create GCS designs that conform to the specified deformations.
Refer to supplementary material for the design parameters.

\begin{figure}[ht]
\centering
\includegraphics[width=0.5\columnwidth]{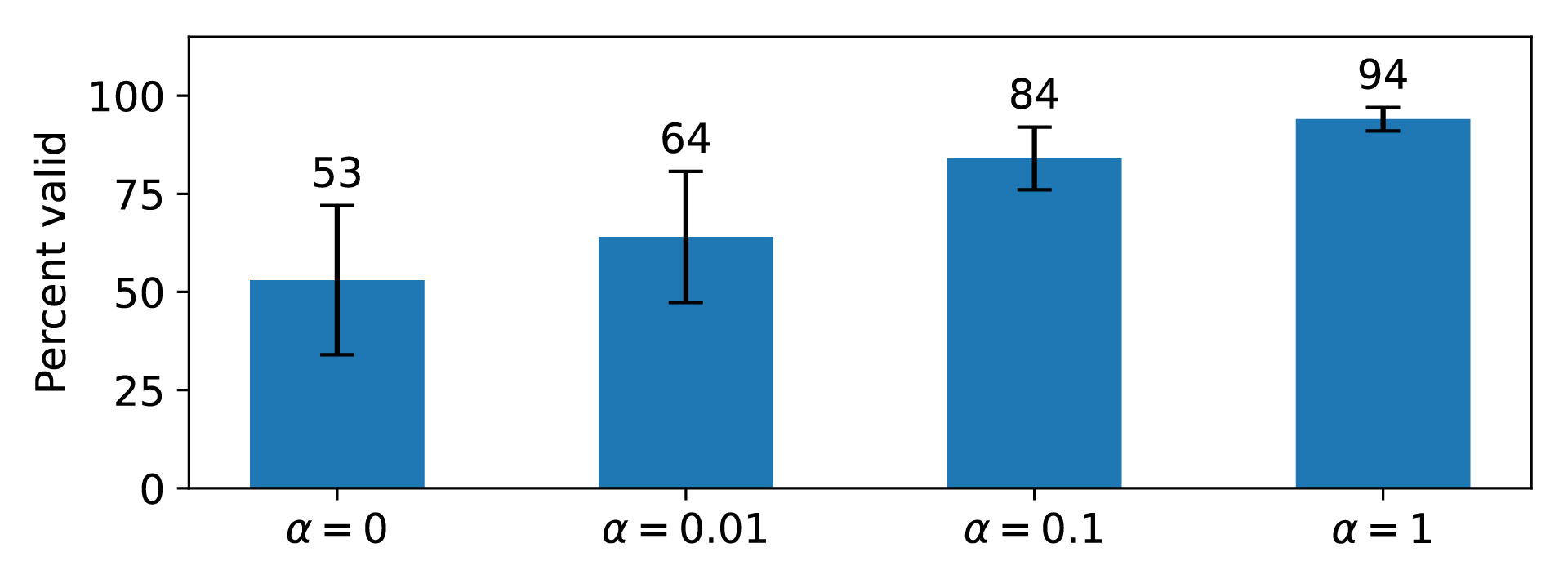}
\caption{\textbf{GCS printability.} The percentage of GCS generated by $\mathcal{I}$ that passes all printability checks when trained with different $\alpha$ values. The error bars depict 95\% confidence intervals.}
\label{fig:alpha}
\end{figure}

We evaluated the printability of generated designs using the two criteria established in our previous work \cite{Snapp2024}:
The base perimeter should be at least 30 mm to provide a substantial contact area with the print bed, and the shell must maintain a minimum distance of 0.01 mm from its center axis to accommodate material deposition.
In Figure~\ref{fig:alpha}, we calculated the percentage of printable predicted designs within the test set for $\mathcal{I}$ trained on different levels of $\alpha$ and observe a positive correlation.
This trend would suggest that further increasing $\alpha$ would continue to improve printability.
However, as $\alpha$ increases, $\mathcal{L}_\mathbf{d}$ dominates $\mathcal{L}_\mathbf{p}$, reverting the training process to the one-to-many mapping problem outlined in Section~\ref{sec:training}.
We did not directly investigate the value of $\alpha$ for which this behavior begins. 

\subsection{Comparison to Alternative Approaches}

We evaluate the TNN against two alternative methods, $k$-nearest neighbors ($k$NN) and FEM, assessing their performance and practicality.

\subsubsection{Nearest Neighbors}

We trained a $k$NN model with $k=1$ for forward and inverse design and evaluated its performance (Table~\ref{tab:knnresults}).
Compared to the TNN, $k$NN is more accurate for stiffness but less accurate for work and maximum displacement, signifying no clear best method concerning these metrics.
However, $k$NN presents several significant limitations.

First, $k$NN has poor scalability.
While the GCS parameterization benefits from a relatively small parameter set, the curse of dimensionality quickly becomes an issue as the parameterization becomes more complex or more materials are added.
From a storage scalability perspective, $k$NN requires storing the entire training dataset, leading to increased model size and longer inference times.

Second, when $k>1$, $k$NN interpolates between the closest designs, lacking a straightforward mechanism to ensure printability.
For instance, interpolating between different one-hot-encoded materials will always produce invalid results.
We have not validated that the nearest design in the normalized parameter space is the most similar.
This leads to a broader question of accurately assessing ``nearness" in non-Euclidean spaces, such as sparse one-hot encoded spaces or other normalized parameter spaces, which warrants its own investigation.

Third, $k$NN cannot transfer knowledge to structures with different parameterizations or performance representations.
Transfer learning reduces experimental data requirements, making it crucial for expanding to other structures and practical applications.

\begin{table}[ht]
    \centering
    \caption{\textbf{Comparison to nearest neighbors.}
    We train a $k$-nearest neighbors ($k$NN) model with $k=1$ and compare its forward and inverse design performance to the TNN.
    For inverse design, we display the results for the $\mathcal{I}$ trained with $\alpha=1$.
    The metrics with higher accuracy for each model are bolded.}
    \label{tab:knnresults}
    \begin{tabular}{llll}
    \toprule
        &  & \multicolumn{1}{c}{\textbf{R}$^2$} & \multicolumn{1}{c}{\textbf{MAE}} \\
    \midrule
    \multicolumn{4}{c}{\textbf{Forward Design}}\\
    \midrule
    \multirow{3}{*}{TNN} & Stiffness (N/mm) & $0.66\pm0.03$ & $250\pm20$ \\
        & \textbf{Work (J)} & $0.96\pm0.002$ & $1.3\pm0.04$ \\
        & \textbf{Max. displacement (mm)} & $0.94\pm0.007$ & $0.50\pm0.03$ \\
    \midrule
    \multirow{3}{*}{$k$NN} & \textbf{Stiffness (N/mm)} & $0.82\pm0.02$ & $172\pm5$ \\
        & Work (J) & $0.92\pm0.004$ & $1.7\pm0.04$ \\
        & Max. displacement (mm) & $0.90\pm0.007$ & $0.60\pm0.01$ \\
    \midrule
    \multicolumn{4}{c}{\textbf{Inverse Design}}\\
    \midrule
    \multirow{3}{*}{TNN} & Stiffness (N/mm) & $0.68\pm0.04$ & $240\pm20$ \\
        & Work (J) & $0.98\pm0.0004$ & $0.91\pm0.08$ \\
        & \textbf{Max. displacement (mm)} & $0.99\pm0.001$ & $0.23\pm0.01$ \\
    \midrule
    \multirow{3}{*}{$k$NN} & \textbf{Stiffness (N/mm)} & $0.92\pm0.01$ & $110\pm6$ \\
        & \textbf{Work (J)} & $0.99\pm0.001$ & $0.62\pm0.01$ \\
        & Max. displacement (mm) & $0.95\pm0.003$ & $0.49\pm0.01$ \\
    \bottomrule
\end{tabular}
\end{table}

\subsubsection{Finite Element Method}

We compared the accuracy and speed of $\mathcal{F}$ to those produced by FEM.
We used Abaqus to numerically derive the compressive force-displacement curve for a GCS design.
Refer to supplementary \S2 for details on the simulation setup.

\begin{figure}[ht]
\centering
\includegraphics[width=0.5\columnwidth]{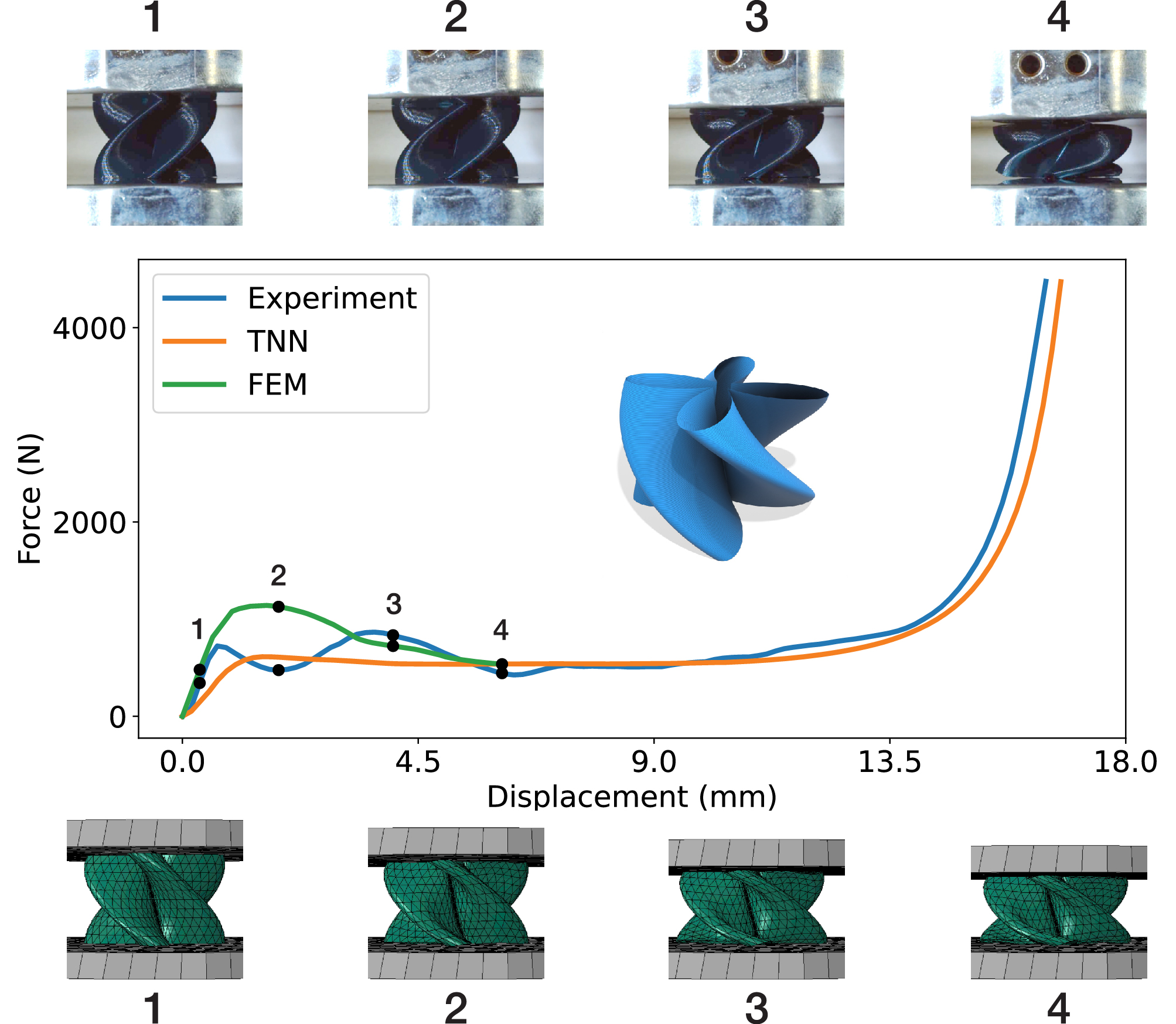}
\caption{\textbf{Comparison to FEM.}
We display the force-displacement curves for a GCS obtained through experimentation (blue), predicted by $\mathcal{F}$ (orange), and simulated via FEM (green).
Four points along the curves indicate the deformations from experimentation (top) and FEM (bottom).}
\label{fig:femresults}
\end{figure}

Figure~\ref{fig:femresults} presents the force-displacement curves obtained from Abaqus and $\mathcal{F}$.
FEM accurately portrayed the linear elastic force-displacement relationship, but the computation begun to lose accuracy for the nonlinear plastic deformations.
The FEM simulation ultimately terminated prematurely and failed to converge beyond a fraction of the total experimental displacement.
In this plastic region of compression, the numerous self-collisions and tearings presented computational challenges for FEM that were not easily addressed.

Experimental results for a GCS design can be obtained in 25 minutes (10 minutes for fabrication and 15 minutes for compression testing).
Our TNN has inference times of $<20$ ms with under one hour to train.
In comparison, the compression test simulation time was 74 minutes on an 8-core CPU with 32 GB of RAM.

We used experimental data instead of simulated data to explore predictive capabilities, as a publicly available dataset capturing the mechanical behavior of interest exists.
However, if such data is not accessible, using synthetic datasets from simulation to train machine learning models is a common strategy, provided that the simulation methods accurately model the behavior of interest.

\subsection{Applications}

We used $\mathcal{I}$ to generate GCS with mechanical behavior tailored to two applications: impact absorption and material emulation.
Refer to supplementary \S3 for more details on each application.

\subsubsection{Impact Absorption}

An impact-absorbing structure must absorb the total impact energy while containing peak forces within specified limits to prevent damage or injury.
Given a force threshold $F$, we optimized for a force-displacement curve meeting (or exceeding) a target energy absorption $E$ to use as input for $\mathcal{I}$,
\begin{equation}
    \label{eq:impact}
    \underset{\text{valid }\mathbf{p}}{\arg\min}\quad E-E_F(\mathbf{p}),
\end{equation}
where $E_F(\mathbf{p})$ denotes the calculated energy absorption (work) before exceeding $F$ (Figure~\hyperref[fig:applications]{\ref*{fig:applications}a}).

We identified GCS optimized for impact absorption in the context of the egg drop test (Figure~\hyperref[fig:applications]{\ref*{fig:applications}b}).
This test involves constructing padding to protect an egg from breaking during a substantial fall.
In our experimental setup, we dropped eggs from 50 cm onto a pad containing four GCS parts.
We set $F=10\text{ N}$ and $E=0.0735\text{ J}$ for the target force-displacement curve.
Using $\mathcal{I}$, we optimized for a GCS design that absorbs the impact energy of the drop without breaking the egg.
We tested three setups, each with five eggs: the optimized pad, an unoptimized pad, and no pad.
We observed a 100\% survival rate for the optimized pad, while the unoptimized pad and no pad showed significantly lower survival rates of 20\% and 0\%, respectively.

\subsubsection{Material Emulation}

\begin{figure}[ht]
\centering
\includegraphics[width=\columnwidth]{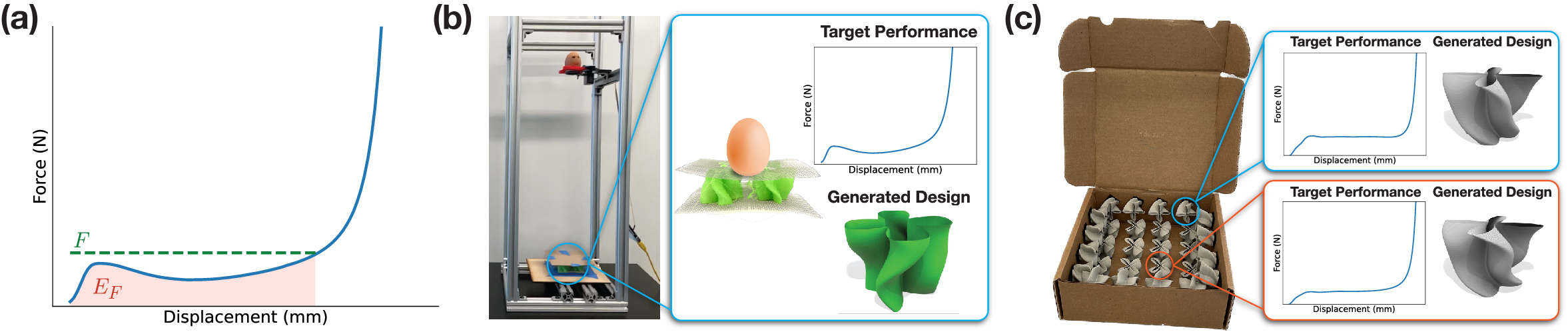}
\caption{\textbf{Applications.}
(\textbf{a}) Given a force threshold $F$, $E_F$ is the energy absorbed before exceeding $F$. 
(\textbf{b}) Custom GCS create padding for the egg drop test. A pad of four GCS absorbs the impact energy of an egg dropped from 50 cm without breaking it.
(\textbf{c}) GCS emulating the mechanical properties of polyurethane foam (PUR), a common material in packaging.}
\label{fig:applications}
\end{figure}

Our TNN enables the creation of GCS that emulate the mechanical behaviors of different materials.
By mimicking the behavior of other materials, one can optimize for non-mechanical properties like weight, cost, and fabrication time.
We designed GCS parts that replicate the behavior of polyurethane (PUR) foam (Figure~\hyperref[fig:gcs]{\ref*{fig:applications}c}), a material commonly employed in packaging.

\section{Discussion}

Discrepancies between predicted and actual curves can come from model prediction errors and lost information from PCA compression.
However, we did not examine which source of error contributes to poor predictions.
In the future, we plan to extend our investigation to look at the performance of PCA compression and explore nonlinear compression methods such as autoencoders.

Real-world constraints often restrict parameters such as height, mass, or material.
However, our current TNN architecture does not allow for user-defined values of generated design parameters, suggesting a clear direction for future enhancements.
One potential strategy is to explore conditioning techniques employed in other neural network architectures \cite{Mirza2014} to grant users fine-grained control over the generated design parameters.

Exploring transfer learning techniques for our TNN presents an exciting avenue for extending its capabilities to diverse 3D printable structures.
Fabricated structures span various parameterizations, yielding structures like lattices \cite{Martinez2019, VantSant2023, Tozoni2020}, crossed barrels \cite{Gongora2020}, and foams \cite{Martinez2016}.
Transferring the acquired design-performance knowledge to different structures, especially those with limited empirical data, holds significant promise for future research.

Finally, understanding how simulation and experimentation can be used in unison to predict high-deformation mechanical properties is essential for future work.
Such approaches offer viable alternatives to collecting extensive experimental datasets, a process typically reliant on access to self-driving labs.
We hope to learn how experimental data can improve simulated outcomes and how much experimental data is needed for this purpose.

\section{Conclusion}

We explore using a Tandem Neural Network (TNN) for the forward and inverse design of FDM 3D printed shells, representing a diverse and versatile class of structures.
Trained on a comprehensive experimental dataset, our TNN reveals the intricate design-performance relationship between shell parameters and compressive behaviors.
By utilizing entire force-displacement curves as performance representations, the network captures a range of nonlinear elastoplastic and hyperelastic deformations.
In forward design, our TNN predicts these nonlinear force-displacement curves based on shell design parameters.
Conversely, in inverse design, the network generates shell designs that exhibit specific desired compressive deformations.
We validate generated shell designs through fabrication and testing. 
Additionally, we generate shells with tailored mechanical properties for several applications.
To encourage further exploration, we make our code and processed dataset publicly available\footnote{\url{https://github.com/samsilverman/nonlinear-deformation-design}}.

\section{Acknowledgements}

The authors thank Adedire Adesiji for brainstorming, assistance in constructing applications, and photography; Helena Gill, Xingjian Han, and Abinit Sati for their work on constructing and running the impact absorption application; and Peter Yichen Chen for his discussions and input.

\section{Author contributions}

Samuel Silverman: Conceptualization, Methodology, Software, Investigation, Writing-Original Draft. 
Kelsey L. Snapp: Conceptualization, Investigation, Validation, Writing-Review \& Editing.
Keith A. Brown: Conceptualization, Methodology, Writing-Review \& Editing, Supervision.
Emily Whiting: Conceptualization, Methodology, Writing-Original Draft, Writing-Review \& Editing, Supervision.

\section{Author Disclosure Statement}

No competing ﬁnancial interests exist.

\section{Funding Information}

This work was supported by the US Army CCDC Soldier Center (contract W911QY2020002). 

\bibliographystyle{vancouver}
\bibliography{main}

\end{document}